\documentstyle[sprocl]{article}

\newenvironment{scalepic}[3]
  {\begin{center} \stepcounter{equation}
\def\@currentlabel{\theequation}      
   \label{#3}
   \begin{picture}(350,50)(-40,-25)%
   \put(0,10){\makebox(0,0)[rb]{#1}}
   \put(0,-10){\makebox(0,0)[rt]{#2}}
   \put(0,0){\makebox(-50,0){$\updownarrow$}}
   \put(0,0){\vector(1,0){300}}
   \put(356,0){\makebox(0,0)
} }
  { \end{picture} \end{center} }
\newcommand{\scaleitem}[3]
 { \put(#1,0){ \put(0,-5){\line(0,1){10}}
               \put(0,10){\makebox(0,0)[b]{#2}}
               \put(0,-10){\makebox(0,0)[t]{#3}} }}

\input{psfig}

\def\Journal#1#2#3#4{{#1} {\bf #2}, #3 (#4)}

\def\NPB{{\em Nucl. Phys.} B}
\def\PLB{{\em Phys. Lett.}  B}
\def\PRL{\em Phys. Rev. Lett.}
\def\PRD{{\em Phys. Rev.} D}

\def\be{\begin{equation}}
\def\ee{\end{equation}}
\def\bea{\begin{eqnarray}}
\def\eea{\end{eqnarray}}
\def\simlt{\stackrel{<}{{}_\sim}}
\def\simgt{\stackrel{>}{{}_\sim}}
\def\draftlabel#1{{\@bsphack\if@filesw {\let\thepage\relax
  \xdef\@gtempa{\write\@auxout{\string
    \newlabel{#1}{{\@currentlabel}{\thepage}}}}}\@gtempa
    \if@nobreak \ifvmode\nobreak\fi\fi\fi\@esphack}
     \gdef\@eqnlabel{#1}}
\def\@eqnlabel{}
\def\@vacuum{}
\def\draftmarginnote#1{\marginpar{\raggedright\scriptsize\tt#1}}
\def\draft{\oddsidemargin -.5truein
        \def\@oddfoot{\sl preliminary draft \hfil
        \rm\thepage\hfil\sl\today\quad\militarytime}
        \let\@evenfoot\@oddfoot \overfullrule 3pt
        \let\label=\draftlabel
        \let\marginnote=\draftmarginnote
   
\def\@eqnnum{(\theequation)\rlap{\kern\marginparsep\tt\@eqnlabel}%
\global\let\@eqnlabel\@vacuum}  }
\def\preprint{\twocolumn\sloppy\flushbottom\parindent 1em
        \leftmargini 2em\leftmarginv .5em\leftmarginvi .5em
        \oddsidemargin -.5in    \evensidemargin -.5in
        \columnsep 15mm \footheight 0pt
        \textwidth 250mmin      \topmargin  -.4in
        \headheight 12pt \topskip .4in
        \textheight 175mm
        \footskip 0pt
        
\def\@oddhead{\thepage\hfil\addtocounter{page}{1}\thepage}
        \let\@evenhead\@oddhead \def\@oddfoot{} \def\@evenfoot{}  }
\def\titlepage{\@restonecolfalse\if@twocolumn\@restonecoltrue\onecolumn
     \else \newpage \fi \thispagestyle{empty}\c@page\z@
        \def\thefootnote{\fnsymbol{footnote}} }
\def\endtitlepage{\if@restonecol\twocolumn \else  \fi
        \def\thefootnote{\arabic{footnote}}
        \setcounter{footnote}{0}}  
\catcode`@=12
\relax

\begin{document}


\begin{flushright}
CPHT-CL741.0999\\
hep-th/9909212
\end{flushright}
\vspace{0.5cm}

\title{MASS SCALES IN STRING AND M-THEORY$^\dagger$}

\author{I. Antoniadis}

\address{Centre de Physique Th{\'e}orique (CNRS UMR 7644),\\ 
Ecole Polytechnique,\\ F-91128 Palaiseau, France,\\ 
E-mail: antoniad@cpht.polytechnique.fr} 


\maketitle\abstracts{\small I review the relations between mass scales
in various string theories and in M-theory. I discuss physical motivations
and possible consistent realizations of large volume compactifications and
low string scale.\\
Large {\it longitudinal} dimensions, seen by Standard Model particles,
imply in general that string theory is strongly coupled unless its tension
is close to the compactification scale. Weakly coupled, low-scale strings
can in turn be realized only in the presence of extra large {\it
transverse} dimensions, seen through gravitational interactions, or in the
presence of infinitesimal string coupling. In the former case, quantum
gravity scale is also low, while in the latter, gravitational and string
interactions remain suppressed by the four-dimensional Planck mass. There
is one exception in this general rule, allowing for large longitudinal
dimensions without low string scale, when Standard Model is embedded in a
six-dimensional fixed-point theory described by a tensionless string.\\
Extra dimensions of size as large as TeV$^{-1}\simeq 10^{-16}$ cm are
motivated from the problem of supersymmetry breaking in string theory,
while TeV scale strings offer a solution to the gauge hierarchy problem,
as an alternative to softly broken supersymmetry or technicolor. I
discuss these problems in the context of the above mentioned
string realizations, as well as the main physical implications both in
particle accelerators and in experiments that measure gravity at
sub-millimeter distances.}

\vspace{0.6cm}
\noindent\footnotesize{$^\dagger$Lectures given at the Spring Workshop
``on Superstrings and Related Matters", ICTP, Trieste, Italy, 22-30 March
1999, and at the Advanced School on ``Supersymmetry in the Theories of
Fields, Strings and Branes", Sandiago de Compostela, Spain, 26-31 July
1999. A short version was given as an invited talk at Strings 99, Potsdam,
Germany, 19-24 July 1999 and at the European Program meeting on ``Quantum
Aspects of Gauge Theories, Supersymmetry and Unification", Paris, France,
1-7 September 1999.}\\ 
\begin{flushleft}
September 1999
\end{flushleft}

\newpage\normalsize
\section{Outline}
The outline of this review article is the following:\\ 

\noindent
1.1~ Preliminaries\\
{\bf 2.}~~ Heterotic string and motivations for large volume
compactifications\\  
2.1~ Gauge coupling unification\\
2.2~ Supersymmetry breaking by compactification\\
{\bf 3.}~~ M-theory on $S^1/Z_2$``$\times$"Calabi-Yau\\
{\bf 4.}~~ Type I/I$^\prime$ string theory and D-branes\\
4.1~ Low-scale strings and extra-large transverse dimensions\\
4.2~ Relation type I/I$^\prime$ -- heterotic\\
{\bf 5.}~~ Type II theories\\
5.1~ Low-scale IIA strings and tiny coupling\\
5.2~ Large dimensions in type IIB\\
5.3~ Relation type II -- heterotic\\
{\bf 6.}~~ Main experimental tests for particle accelerators\\
6.1~ Longitudinal dimensions\\
6.2~ Transverse dimensions and low-scale quantum gravity\\
6.3~ Low-scale strings\\
{\bf 7.}~~ Physics\\
7.1~ Gauge hierarchy\\
7.2~ Unification\\
7.3~ Supersymmetry breaking\\
{\bf 8.}~~ Gravity modification and sub-millimeter forces\\
References

\subsection{Preliminaries}
In critical (ten) dimensions, any consistent superstring theory has two
parameters: a mass (or length) scale $M_s$ ($l_s=M_s^{-1}$), and a
dimensionless string coupling $\lambda_s$ given by the vacuum expectation
value (VEV) of the dilaton field
$e^{<\phi>}=\lambda_s$~\cite{strings,sao}: 
\be
D=10:\qquad\quad M_s=l_s^{-1}\qquad \lambda_s\, .
\label{tenD}
\ee
Upon compactification in $D=4$ dimensions on a compact manifold of volume
$V$, these parameters determine the four-dimensional (4d) Planck mass (or
length) $M_p$ ($l_p=M_p^{-1}$) and the dimensionless gauge coupling $g$
at the string scale. For simplicity, in the following we drop all
numerical factors from our formulae, while, when needed, we use the
numerical values:
\be
D=4:\qquad\quad M_p\simeq 1.2\times 10^{19}\ {\rm GeV}\qquad 
g\simeq 1/5\, .
\label{fourD}
\ee
Moreover, the weakly coupled condition implies that $\lambda_s<<1$. Our
method in the following consists in expressing the 10d parameters
$(M_s,\lambda_s)$ in terms of the 4d ones and the compactification
volume, in heterotic ($s=H$), type I ($s=I$) and type II ($s=II$) string
theories, and then discuss the conditions on possible large volume or low
string scale realizations, keeping the string coupling small.

An important point is that the compactification volume will always be
chosen to be bigger than unity in string units, $V>l_s^6$. This can
be done by a T-duality transformation which exchanges the role of
the Kaluza-Klein (KK) momenta $p$ with the string winding modes $w$. For
instance, in the case of one compact dimension on a circle of radius $R$,
they read:
\be
p={m\over R} \qquad ;\qquad w={nR\over l_s^2}\, ,
\label{pw}
\ee
with integers $m,n$. T-duality inverts the compactification radius and
rescales the string coupling:
\be
R\to{l_s^2\over R}\qquad \lambda_s\to\lambda_s{l_s\over R}\, ,
\label{Tdual}
\ee
so that the lower-dimensional coupling $\lambda_s\sqrt{l_s/R}$ remains
invariant. When $R$ is smaller than the string scale, the winding modes
become very light, while T-duality trades them as KK momenta in terms of
the dual radius ${\tilde R}\equiv l_s^2/R$. The enhancement of the string
coupling is then due to their multiplicity which diverges in the limit
$R\to 0$ (or ${\tilde R}\to\infty$).

\section{Heterotic string and motivations for large volume
compactifications}

In heterotic string, gauge and gravitational interactions appear at the
same (tree) level of perturbation theory (spherical world-sheet
topology), and the corresponding effective action is~\cite{strings,sao}:
\be
S=\int d^4x{V\over\lambda_H^2}(l_H^{-8}{\cal R}+l_H^{-6}F^2)\, ,
\ee
upon compactification in four dimensions. Here, for simplicity, we kept
only the gravitational and gauge kinetic terms, in a self-explanatory
notation. Identifying their respective coefficients with the 4d
parameters $1/l_p^2$ and $1/g^2$, one obtains:
\be
M_H=gM_p\qquad\qquad \lambda_H=g{\sqrt{V}\over l_H^3}\, .
\label{het}
\ee
Using the values (\ref{fourD}), one obtains that the heterotic string
scale is near the Planck mass, $M_H\simeq 10^{18}$, while the string is
weakly coupled when the internal volume is of order of the string scale,
$V\sim l_s^6$. However, despite this fact, there are physical motivations
which suggest that large volume compactifications, and thus strong
coupling, may be relevant in physics~\cite{ia}. These come from gauge
coupling unification and supersymmetry breaking by compactification,
which we discuss below.

\subsection{Gauge coupling unification}

It is a known fact that the three gauge couplings of the Standard Model,
when extrapolated at high energies assuming the particle content of its
$N=1$ minimal supersymmetric extension (MSSM), they meet at an energy
scale $M_{\rm GUT}\simeq 2\times 10^{16}$ GeV. At the one-loop level, one
has:
\be
{1\over g^2_a(\mu)}={1\over g^2}+{b_a\over 4\pi}
\ln{M_{\rm GUT}^2\over\mu^2}\, ,
\ee
where $\mu$ is the energy scale and $a$ denotes the 3 gauge group factors
of the Standard Model $SU(3)\times SU(2)\times U(1)$. The value of
$M_{\rm GUT}$ is very near the heterotic string scale, but it differs by
roughly two orders of magnitude. If one takes seriously this discrepancy,
a possible way to explain it is by introducing large compactification
volume.

Consider for instance one large dimension of size $R$, so that 
$V\sim Rl_H^5$. Identifying $M_{\rm GUT}$ with the compactification scale
$R^{-1}$, this requires $R\sim 100l_H$. Alternatively, one can use string
threshold corrections which grow linearly with $R$~\cite{dkl}. Assuming
that they can account for the discrepancy, one needs roughly 
$R/l_H\sim\ln(M_H^2/M_{\rm GUT}^2)\sim 10$. As a result, the string
coupling (\ref{het}) equals $\lambda_H\sim 0.5 - 2$ which enters in the
strongly coupled regime.

\subsection{Supersymmetry breaking by compactification}

In contrast to ordinary supergravity, where supersymmetry breaking can be
introduced at an arbitrary scale, through for instance the gravitino,
gaugini and other soft masses, in string theory this is not possible
(perturbatively). The only way to break supersymmetry at a scale
hierarchically smaller than the (heterotic) string scale is by
introducing a large compactification radius whose size is set by the
breaking scale. This has to be therefore of the order of a few TeV in
order to protect the gauge hierarchy. An explicit proof exists for
toroidal and fermionic constructions, although the result is believed to
apply to all compactifications~\cite{ablt,kp}. This is one of the very
few general predictions of perturbative (heterotic) string theory that
leads to the spectacular prediction of the possible existence of extra
dimensions accessible to future accelerators~\cite{ia}. The main
theoretical problem is though the strong coupling, as mentioned above.

The strong coupling problem can be understood from the effective field
theory point of view from the fact that at energies higher than the
compactification scale, the KK excitations of gauge bosons and other
Standard Model particles will start being produced and contribute to
various physical amplitudes. Their multiplicity turns very rapidly the
logarithmic evolution of gauge couplings into a power
dependence~\cite{tv}, invalidating the perturbative description, as
expected in a higher dimensional non-renormalizable gauge theory. A
possible way to avoid this problem is to impose conditions which prevent
the power corrections to low-energy couplings~\cite{ia}. For gauge
couplings, this implies the vanishing of the corresponding
$\beta$-functions, which is the case for instance when the KK modes are
organized in multiplets of $N=4$ supersymmetry, containing for every
massive spin-1 excitation, 2 Dirac fermions and 6 scalars. Examples of
such models are provided by orbifolds with no $N=2$ sectors with respect
to the large compact coordinate(s).

The simplest example of a one-dimensional orbifold is an interval of
length $\pi R$, or equivalently $S^1/Z_2$ with $Z_2$ the coordinate
inversion. The Hilbert space is composed of the untwisted sector,
obtained by the $Z_2$-projection of the toroidal states (\ref{pw}), and
of the twisted sector which is localized at the two end-points of the
interval, fixed under the $Z_2$ transformations. This sector is chiral
and can thus naturally contain quarks and leptons, while gauge fields
propagate in the (5d) bulk.

Similar conditions should be imposed to Yukawa's and in principle to
higher (non-renormalizable) effective couplings in order to ensure a soft
ultraviolet (UV) behavior above the compactification scale. We now know
that the problem of strong coupling can be addressed using string
S-dualities which invert the string coupling and relate a strongly
coupled theory with a weakly coupled one~\cite{sao}. For instance, as we
will discuss below, the strongly coupled heterotic theory with one large
dimension is described by a weakly coupled type IIB theory with a tension
at intermediate energies $(Rl_H)^{-1/2}\simeq 10^{11}$ GeV~\cite{ap}.
Furthermore, non-abelian gauge interactions emerge from tensionless
strings~\cite{w95} whose effective theory describes a higher-dimensional
non-trivial infrared fixed point of the renormalization group~\cite{sei}.
This theory incorporates all conditions to low-energy couplings that
guarantee a smooth UV behavior above the compactification scale. In
particular, one recovers that KK modes of gauge bosons form $N=4$
supermultiplets, while matter fields are localized in four dimensions. It
is remarkable that the main features of these models were captured
already in the context of the heterotic string despite its strong
coupling~\cite{ia}.

In the case of two or more large dimensions, the strongly coupled
heterotic string is described by a weakly coupled type IIA or type
I/I$^\prime$ theory~\cite{ap}. Moreover, the tension of the dual string
becomes of the order or even lower than the compactification scale. In
fact, as it will become clear in the following, in the context of any
string theory other than the heterotic, the simple relation (\ref{het})
that fixes the string scale in terms of the Planck mass does not hold and
therefore the string tension becomes an arbitrary parameter~\cite{w}. It
can be anywhere below the Planck scale and as low as a few TeV~\cite{l}.
The main advantage of having the string tension at the TeV, besides its
obvious experimental interest, is that it offers an automatic solution to
the problem of gauge hierarchy, alternative to low-energy supersymmetry or
technicolor~\cite{add,aadd,ab}.

\section{M-theory on $S^1/Z_2$``$\times$"Calabi-Yau}

The strongly coupled $E_8\times E_8$ heterotic string compactified on a
Calabi-Yau manifold (CY) of volume $V$ is described by the 11d M-theory
compactified on an interval $S^1/Z_2$ of length $\pi R_{11}$ times the
same Calabi-Yau~\cite{hw}. Gravity propagates in the 11d bulk, containing
besides the metric and the gravitino a 3-form potential, while gauge
interactions are confined on two 10d boundaries (9-branes) localized at
the two end-points of the interval and containing one $E_8$ factor each.
The corresponding effective action is
\be
S_H=\int d^4x V({R_{11}\over l_M^9}{\cal R}+{1\over l_M^6}F^2)\, .
\label{SH}
\ee
It follows that
\be
l_M=(g^2V)^{1/6}\qquad\qquad R_{11}=g^2{l_M^3\over l_P^2}\, .
\label{Mth}
\ee
The validity of the 11d supergravity regime is when $R_{11}>l_M$ and
$V>l_M^6$ implying $g<1$ by virtue of eq.(\ref{Mth}). Comparison with
the heterotic relations (\ref{het}) yields:
\be
l_M=l_H\lambda_H^{1/3}\qquad\qquad R_{11}=l_H\lambda_H\, ,
\label{M-het}
\ee
which shows in particular that $R_{11}$ is the string coupling in
heterotic units. As a result, at strong coupling $\lambda_H>1$ the M
theory scale and the 11d radius are larger than the heterotic length:
$R_{11}>l_M>l_H$.

Imposing the M-theory scale $l_M^{-1}$ to be at 1 TeV, one finds from the
relations (\ref{Mth}) a value for the radius of the 11th dimension of the
size of the solar system, $R_{11}\simeq 10^8$ kms, which is obviously
excluded experimentally. On the other hand, imposing a value for
$R_{11}\simeq 1$ mm which is the shortest length scale that gravity is
tested experimentally, one finds a lower bound for the M-theory scale
$l_M^{-1}\simgt 10^7$ GeV~\cite{ckm}.

While the relations (\ref{Mth}) seem to impose no theoretical constraint
to $l_M$, there is however another condition to be imposed beyond the
classical approximation~\cite{w}. This is because at the next order the
factorized space $S^1/Z_2\times\rm CY$ is not any more solution of the
11d supergravity equations, which require the size of the Calabi-Yau
manifold to depend on the 11th coordinate $x_{11}$ along the interval.
This can be seen for instance from the supersymmetry transformation of
the 3-form potential (with field-strength $G^{(4)}$) which acquires non
vanishing contributions from the 10d boundaries:
\be
\delta G^{(4)}=l_M^6\delta(x_{11})\left({\rm tr}F\wedge F-
{1\over 2}{\rm tr}{\cal R}\wedge{\cal R}\right)+
\left( x_{11}\leftrightarrow \pi R_{11}-x_{11}, 
F\leftrightarrow F'\right)\, .
\ee
As a result, the volume of CY varies linearly along the interval, to
leading order:
\be
V(x_{11})=V(0)-x_{11}l_M^3 \int_{\rm CY} \omega\wedge\left(
{\rm tr}F'\wedge F'- {\rm tr}F\wedge F\right) \, ,
\label{V}
\ee
where $\omega\sim V^{1/3}$ is the K\"ahler form on the six-manifold CY.

It follows that there is an upper bound on $R_{11}$, otherwise the gauge
coupling in one of the two walls blows up when the volume of CY shrinks
to zero size. Choosing $V(0)\equiv V$ and imposing $V(\pi R)\ge 0$, 
eq.(\ref{V}) yields $R_{11}\simlt V^{2/3}/l_M^3$ and through the
relations (\ref{Mth}):
\be
l_P\simgt g^{5/3}l_M=g^2 V^{1/6}\, .
\ee
This implies a lower bound for the M-theory scale 
$l_M^{-1}\simgt g^{5/3}M_P$, or equivalently for the unification scale
$M_{\rm GUT}\equiv V^{-1/6}\simgt g^2M_P$. Taking into account the
numerical factors, on finds for the lower bound the right order of
magnitude $M_{\rm GUT}\sim 10^{16}$ GeV, providing a solution to the
perturbative discrepancy between the unification and heterotic string
scales, discussed in section {\it 2.1}~\cite{w}. Note that this bound
does not hold in the case of symmetric embedding, where one has 
${\rm tr}F'\wedge F'-{\rm tr}F\wedge F=0$ and thus the correction in
eq.(\ref{V}) vanishes.

\section{Type I/I$^\prime$ string theory and D-branes}

In ten dimensions, the strongly coupled $SO(32)$ heterotic string is 
described by the type I string, or upon T-dualities to type
I$^\prime$~\cite{pw,sao}.\footnote{In lower dimensions, type I$^\prime$
theories can also describe a class of M-theory compactifications.} Type
I/I$^\prime$ is a theory of closed and open unoriented strings. Closed
strings describe gravity, while gauge interactions are described by open
strings whose ends are confined to propagate on D-branes. It follows that
the 6 internal compact dimensions are separated into longitudinal
(parallel) and transverse to the D-branes. Assuming that the Standard
Model is localized on a $p$-brane with $p\ge 3$, there are $p-3$
longitudinal and $9-p$ transverse compact dimensions. In contrast to the
heterotic string, gauge and  gravitational interactions appear at
different order in perturbation theory and the corresponding effective
action reads~\cite{strings,sao}:
\be
S_{I}=\int d^{10}x \frac{1}{\lambda_I^2 l_I^8} {\cal R} + 
\int d^{p+1}x \frac{1}{\lambda_I l_I^{p-3}} F^2\, ,
\label{SI}
\ee
where the $1/\lambda_I$ factor in the gauge kinetic terms corresponds to
the disk diagram. 

Upon compactification in four dimensions, the Planck length and gauge
couplings are given to leading order by
\begin{equation}
\frac{1}{l_P^2}=\frac{V_\parallel V_\perp}{\lambda_I^2 l_I^8}\ ,\qquad
\frac{1}{g^2}=\frac{V_\parallel}{\lambda_I l_I^{p-3}}\, ,
\label{I}
\end{equation}
where $V_\parallel$ ($V_\perp$) denotes the compactification volume 
longitudinal (transverse) to the $p$-brane. From the second relation
above, it follows that the requirement of weak coupling $\lambda_I<1$
implies that the size of the longitudinal space must be of order of the
string length ($V_\parallel\sim l_I^{p-3}$), while the transverse volume
$V_\perp$ remains unrestricted. One thus has
\begin{equation}
M_P^2=\frac{1}{g^4 v_\parallel}M_I^{2+n}R_\perp^n\ ,\qquad
\lambda_I =g^2 v_\parallel\, ,
\label{treei}
\end{equation}
to be compared with the heterotic relations (\ref{het}). Here, 
$v_\parallel\simgt 1$ is the longitudinal volume in string units, 
and we assumed an isotropic transverse space of $n=9-p$ compact 
dimensions of radius $R_\perp$.

\subsection{Low-scale strings and extra-large transverse dimensions}

From the relations (\ref{treei}), it follows that the type I/I$^\prime$
string scale can be made hierarchically smaller than the Planck mass at
the expense of introducing extra large transverse dimensions that
interact only gravitationally, while keeping the string coupling
weak~\cite{aadd,st}. The weakness of 4d gravity $M_I/M_P$ is then
attributed to the largeness of the transverse space $R_\perp/l_I$. An
important property of these models is that gravity becomes strong at the
string scale, although the string coupling remains weak. In fact, the
first relation of eq.(\ref{treei}) can be understood as a consequence of
the
$(4+n)$-dimensional Gauss law for gravity, with
\be
G_N^{(4+n)}=g^4 l_I^{2+n}v_\parallel
\label{GN}
\ee
the Newton's constant in $4+n$ dimensions.

To be more explicit, taking the type I string scale $M_I$ to be at 1 TeV,
one finds a size for the transverse dimensions $R_\perp$ varying from
$10^8$ km, .1 mm (10$^{-3}$ eV), down  to .1 fermi (10 MeV) for $n=1,2$,
or 6 large dimensions, respectively. The case $n=1$ corresponds to
M-theory and is obviously experimentally excluded. On the other hand, all
other possibilities are consistent with observations, although barely in
the case $n=2$~\cite{add2}. In particular, sub-millimeter transverse
directions are compatible with  the present constraints from
short-distance gravity measurements which tested Newton's law up to the
cm~\cite{price}. The strongest bounds come from astrophysics and
cosmology and concern mainly the case $n=2$. In fact, graviton emission
during supernovae cooling restricts the 6d Planck scale to be larger than
about 50 TeV, implying $M_I\simgt 7$ TeV, while the graviton decay
contribution to the cosmic diffuse gamma radiation gives even stronger
bounds of about 110 TeV and 15 TeV for the two scales, respectively.

If our brane world is supersymmetric, which protects the hierarchy in the
usual way, the string scale is an arbitrary parameter and can be at
higher energies, in principle up to the Planck scale. However, in the
context of type I/I$^\prime$ theory, the string scale should not be
higher than intermediate energies $M_I\simlt 10^{11}$ GeV, due to the
generic existence of other branes with non supersymmetric world
volumes~\cite{li}. Indeed, in this case, our world would feel the effects
of supersymmetry breaking through gravitationally suppressed interactions
of order $M_I^2/M_P$, that should be less than a TeV. In this context, the
value $M_I\sim 10^{11}$ GeV could be favored, since it would coincide with
the scale of supersymmetry breaking in a hidden sector, without need of
non-perturbative effects such as gaugino condensation. Moreover, the gauge
hierarchy would be minimized, since one needs to introduce transverse
dimensions with size just two orders of magnitude larger than
$l_I$ (in the case of $n=6$) to account for the ratio
$M_I/M_P\simeq 10^{-8}$, according to eq.(\ref{treei}). Note also that
the weak scale $M_W\sim M_I^2/M_P$ becomes T-dual to the Planck scale.

\subsection{Relation type I/I$^\prime$ -- heterotic}

We will now show that the above type I/I$^\prime$ models describe
particular strongly coupled heterotic vacua with large
dimensions~\cite{aq,ap}. More precisely, we will consider the heterotic
string compactified on a 6d manifold with $k$ large dimensions of radius 
$R\gg l_H$ and $6-k$ string-size dimensions and show that for $k\ge 4$ it
has a perturbative type I$^\prime$ description~\cite{ap}. 

In ten dimensions, heterotic and type I theories are related by an
S-duality:
\be
\lambda_I={1\over\lambda_H}\qquad\qquad l_I=\lambda_H^{1/2}l_H\, ,
\label{het-I}
\ee
which can be obtained for instance by comparing eqs.(\ref{het}) with
eqs.(\ref{I}) in the case of 9-branes ($p=9$, $V_\perp=1$,
$V_\parallel=V$). Using from eq.(\ref{het}) that 
$\lambda_H\sim (R/l_H)^{k/2}$, one finds
\be
\lambda_I\sim\left({R\over l_H}\right)^{-k/2}\qquad\qquad
l_I\sim\left({R\over l_H}\right)^{k/4}l_H\, .
\ee
It follows that the type I scale $M_I$ appears as a non-perturbative
threshold in the heterotic string at energies much lower than
$M_H$~\cite{ckm}. For $k<4$, it appears at intermediate energies
$R^{-1}<M_I<M_H$, for $k=4$, it becomes of the order of the
compactification scale $M_I\sim R^{-1}$, while for $k>4$, it appears at
low energies $M_I<R^{-1}$~\cite{aq}. Moreover, since $\lambda_I\ll 1$,
one would naively think that weakly coupled type I theory could describe
the heterotic string with any number $k\ge 1$ of large dimensions.
However, this is not true because there are always some dimensions
smaller than the type I size ($6-k$ for $k<4$ and 6 for $k>4$) and one
has to perform T-dualities (\ref{Tdual}) in order to account for the
multiplicity of light winding modes in the closed string sector, as we
discussed in section {\it 1.1}. Note that open strings have no winding
modes along longitudinal dimensions and no KK momenta along transverse
directions. The T-dualities have two effects: (i) they transform the
corresponding longitudinal directions to transverse ones by exchanging KK
momenta with winding modes, and (ii) they increase the string coupling
according to eq.(\ref{Tdual}) and therefore it is not clear that type
I$^\prime$ theory remains weakly coupled.

Indeed for $k<4$, after performing $6-k$ T-dualities on the heterotic
size dimensions, with respect to the type I scale, one obtains a type
I$^\prime$ theory with D($3+k$)-branes but strong coupling:
\be
l_H\to{\tilde l}_H\!=\!{l_I^2\over l_H}\!\sim\!
\left({R\over l_H}\right)^{k/2}l_H
\qquad \lambda_I\to{\tilde\lambda}_I\!=\!
\lambda_I\left({l_I\over l_H}\right)^{6-k}\!\sim\!
\left({R\over l_H}\right)^{k(4-k)/4}\!\!\gg\!\! 1\, .
\label{kl4}
\ee
For $k\ge 4$, we must perform T-dualities in all six internal
directions.\footnote{The case $k=4$ can be treated in the same way, since
there are 4 dimensions that have type I string size and remain inert
under T-duality.} As a result, the type I$^\prime$ theory has D3-branes
with $6-k$ transverse dimensions of radius ${\tilde l}_H$ given in
eq.(\ref{kl4}) and $k$ transverse dimensions of radius 
${\tilde R}=l_I^2/R\sim (R/l_H)^{k/2-1}$, while its coupling remains weak
(of order unity):
\be
\lambda_I\to{\tilde\lambda}_I=\lambda_I
\left({l_I\over l_H}\right)^{6-k}\left({l_I\over R}\right)^k\sim 1\, .
\ee

It follows that the type I$^\prime$ theory with $n$ extra-large
transverse dimensions offers a weakly coupled dual description for the
heterotic string with $k=4,5,6$ large dimensions~\cite{ap}. $k=4$ is
described by $n=2$, $k=6$ (for $SO(32)$ gauge group) is described by
$n=6$, while for $n=5$ one finds a type I$^\prime$ model with 5 large
transverse dimensions and one extra-large. The case $k=4$ is particularly
interesting: the heterotic string with 4 large dimensions, say at a TeV,
is described by a perturbative type I$^\prime$ theory with the string
scale at the TeV and 2 transverse dimensions of millimeter size that are
T-dual to the 2 heterotic string size coordinates. This is depicted in
the following diagram, together with the case $k=6$, where we use
heterotic length units $l_H=1$:
\begin{scalepic}
{H: $k=4$}{I$^\prime$: $n=2$}{scal04}
\scaleitem{30}{$l_H$, $R_{5,6}$}{1}
\scaleitem{140}{$R_{1,2,3,4}=R$}{$l_I$}
\scaleitem{250}{$R^2$}{$\tilde R_{5,6}$}
\end{scalepic}
\begin{scalepic}
{H: $k=6$}{I$^\prime$: $n=6$}{scal06}
\scaleitem{30}{$l_H$}{1}
\scaleitem{140}{$R_{1,\cdots,6}=R$}{}
\scaleitem{195}{$R^{3/2}$}{$l_I$}
\scaleitem{250}{$R^2$}{$\tilde R_{1,\cdots,6}$}
\end{scalepic}

\section{Type II theories}

Upon compactification to 6 dimensions or lower, the heterotic string
admits another dual description in terms of type II (IIA or IIB) string
theory~\cite{ht,sao}. Since in 10 dimensions type II theories have $N=2$
supersymmetry,\footnote{Type IIA (IIB) has two 10d supercharges of
opposite (same) chirality.} in contrast to the heterotic string which has
$N=1$, the compactification manifolds on the two sides should be
different, so that the resulting theories in lower dimensions have the
same number of supersymmetries. The first example arises in 6 dimensions,
where the $E_8\times E_8$ heterotic string compactified on the four-torus
$T^4$ is S-dual to type IIA compactified on the $K3$ manifold that has
$SU(2)$ holonomy and breaks half of the supersymmetries. In lower
dimensions, type IIA and type IIB are related by T-duality (or mirror
symmetry). 

Here, for simplicity, we shall restrict ourselves to 4d compactifications
of type II on $K3\times T^2$, yielding $N=4$ supersymmetry, or more
generally on Calabi-Yau manifolds that are $K3$ fibrations, yielding
$N=2$ supersymmetry. They are obtained by replacing $T^2$ by a ``base"
two-sphere over which $K3$ varies, and they are dual to corresponding
heterotic compactifications on $K3\times T^2$. More interesting
phenomenological models with $N=1$ supersymmetry can be obtained by a
freely acting orbifold on the two sides, although the most general $N=1$
compactification would require F-theory on Calabi-Yau fourfolds, which
is poorly understood at present~\cite{pm}. 

In contrast to  heterotic and type I strings, non-abelian gauge symmetries
in type II models arise non-perturbatively (even though at arbitrarily
weak coupling) in singular compactifications, where the massless gauge
bosons are provided by D2-branes in type IIA (D3-branes in IIB) wrapped
around non-trivial vanishing 2-cycles (3-cycles). The resulting gauge
interactions are localized on $K3$ (similar to a Neveu-Schwarz
five-brane), while matter multiplets would arise from further
singularities, localized completely on the 6d internal space~\cite{kv}.

\subsection{Low-scale IIA strings and tiny coupling}

In type IIA non-abelian gauge symmetries arise in six dimensions from
D2-branes wrapped around non-trivial vanishing 2-cycles of a singular
$K3$.\footnote{Note though that the abelian Cartan subgroup is already in
the perturbative spectrum of the Ramond-Ramond sector.} It follows that
gauge kinetic terms are independent of the string coupling
$\lambda_{IIA}$ and the corresponding effective action is~\cite{sao}:
\be
S_{IIA}=\int d^{10}x \frac{1}{\lambda_{IIA}^2 l_{IIA}^8} {\cal R} + 
\int d^6 x {1\over l_{IIA}^2} F^2\, ,
\label{SIIA}
\ee
which should be compared with (\ref{SH}) of heterotic and (\ref{SI}) of
type I/I$^\prime$. As a result, upon compactification in four dimensions,
for instance on a two-torus $T^2$, the gauge couplings are determined by
its size $V_{T^2}$, while the Planck mass is controlled by the 6d string
coupling $\lambda_{6IIA}$:
\be
\frac{1}{g^2}={V_{T^2}\over l_{IIA}^2} \qquad\qquad
\frac{1}{l_P^2}=\frac{V_{T_2}}{\lambda_{6IIA}^2 l_{IIA}^4}
={1\over\lambda_{6IIA}^2}{1\over g^2l_{IIA}^2}\, .
\label{IIA}
\ee

The area of $T^2$ should therefore be of order $l_{IIA}^2$, while the
string scale is expressed by
\be
M_{IIA}=g\lambda_{6IIA}M_P=
g\lambda_{IIA}M_P{l_{IIA}^2\over\sqrt{V_{K3}}}\, ,
\label{IIA2}
\ee
with $V_{K3}$ the volume of $K3$. Thus, in contrast to the type I
relation (\ref{treei}) where only the volume of the internal six-manifold
appears, we now have the freedom to use both the string coupling and
the $K3$ volume to separate the Planck mass from a string scale, say,
at 1 TeV~\cite{l,ap}. In particular, we can choose a string-size internal
manifold, and have an ultra-weak coupling $\lambda_{IIA}=10^{-14}$ to
account for the hierarchy between the electroweak and the Planck
scales~\cite{ap}. As a result, despite the fact that the string scale is
so low, gravity remains weak up to the Planck scale and string
interactions are suppressed by the tiny string coupling, or equivalently
by the 4d Planck mass. Thus, there are no observable effects in particle
accelerators, other than the production of KK excitations along the two
TeV dimensions of $T^2$ with gauge interactions. Furthermore, the
excitations of gauge multiplets have $N=4$ supersymmetry, even when
$K3\times T^2$ is replaced by a Calabi-Yau threefold which is a $K3$
fibration, while matter multiplets are localized on the base (replacing
the $T^2$) and have no KK excitations, as the twisted states of heterotic
orbifolds.

Above, we discussed the simplest case of type II compactifications with
string scale at the TeV and all internal radii having the string size. In
principle, one can allow some of the $K3$ (transverse) directions to be
large, keeping the string scale low. From eq.(\ref{IIA2}), it follows
that the string coupling $\lambda_{IIA}$ increases making gravity strong
at distances $l_P\sqrt{V_{K3}}/l_{IIA}^2$ larger than the Planck length.
In particular, it becomes strong at the string scale (TeV), when
$\lambda_{IIA}$ is of order unity. This corresponds to 
$V_{K3}/l_{IIA}^4\sim 10^{28}$, implying a fermi size for the four $K3$
compact dimensions.

\subsection{Large dimensions in type IIB}

Above we assumed that both directions of $T^2$ have the string size, so
that its volume is of order $l_{IIA}^2$, as implied by eq.(\ref{IIA}).
However, one could choose one direction much bigger than the string scale
and the other much smaller. For instance, in the case of a rectangular
torus of radii $r$ and $R$, $V_{T^2}=rR\sim l_{IIA}^2$ with
$r\gg l_{IIA}\gg R$. This can be treated by performing a T-duality
(\ref{Tdual}) along $R$ to type IIB: $R\to l_{IIA}^2/R$ and
$\lambda_{IIA}\to\lambda_{IIB}=\lambda_{IIA}l_{IIA}/R$ with
$l_{IIA}=l_{IIB}$. One thus obtains:
\be
\frac{1}{g^2}={r\over R} \qquad\qquad
\frac{1}{l_P^2}=\frac{V_{T_2}}{\lambda_{6IIB}^2 l_{IIB}^4}
={R^2\over\lambda_{6IIB}^2}{1\over g^2l_{IIB}^4}\, .
\label{IIB}
\ee
which shows that the gauge couplings are now determined by the ratio of
the two radii, or in general by the shape of $T^2$, while the Planck mass
is controlled by its size, as well as by the 6d type IIB string coupling.
The string scale can thus be expressed as~\cite{ap}:
\be
M_{IIB}^2=g\lambda_{6IIB}{M_P\over R}\, .
\label{IIB2}
\ee

Comparing these relations with eqs.(\ref{IIA}) and (\ref{IIA2}), it is
clear that the situation in type IIB is the same as in type IIA, unless
the size of $T^2$ is much larger than the string length, $R\gg l_{IIB}$.
Since $T^2$ is felt by gauge interactions, its size cannot be larger than
${\cal O}({\rm TeV}^{-1})$ implying that the type IIB string scale should
be much larger than TeV. From eq.(\ref{IIB2}) and $\lambda_{6IIB}<1$, one
finds $M_{IIB}\simlt\sqrt{M_P/R}$, so that the largest value for the
string tension, when $R\sim 1{\rm TeV}^{-1}$, is an intermediate scale
$\sim 10^{11}$ GeV when the string coupling is of order unity. 

As we will show below, this is precisely the case that describes the
heterotic string with one TeV dimension, which we discussed is section 2.
It is the only example of longitudinal dimensions larger than the string
length in a weakly coupled theory. In the energy range between the KK
scale $1/R$ and the type IIB string scale, one has an effective 6d theory
without gravity at a non-trivial superconformal fixed point described by
a tensionless string~\cite{w95,sei}. This is because in type IIB gauge
symmetries still arise non-perturbatively from vanishing 2-cycles of
$K3$, but take the form of tensionless strings in 6 dimensions, given by
D3-branes wrapped on the vanishing cycles. Only after further
compactification does this theory reduce to a standard gauge theory,
whose coupling involves the shape rather than the volume of the
two-torus, as described above. Since the type IIB coupling is of order
unity, gravity becomes strong at the type IIB string scale and the main
experimental signals at TeV energies are similar to those of type IIA
models with tiny string coupling.

\subsection{Relation type II -- heterotic}

We will now show that the above low-scale type II models describe some
strongly coupled heterotic vacua and, in particular, the cases with
$k=1,2,3$ large dimensions that have not a perturbative description in
terms of type I$^\prime$ theory~\cite{ap}. As we described in the
beginning of section 5, in 6 dimensions the heterotic $E_8\times E_8$
superstring compactified on $T^4$ is S-dual to type IIA compactified on
$K3$:
\be
\lambda_{6IIA}={1\over\lambda_{6H}}\qquad\qquad
l_{IIA}=\lambda_{6H}l_H\, ,
\label{het-II}
\ee
which can be obtained, for instance, by comparing eqs.(\ref{IIA}) with
(\ref{het}), using $\lambda_{6H}=\lambda_H l_H^2/\sqrt{V_{T^4}}$.
However, in contrast to the case of heterotic -- type I/I$^\prime$
duality, the compactification manifolds on the two sides are not the same
and a more detailed analysis is needed to study the precise mapping of
$T^4$ to $K3$, besides the general relations (\ref{het-II}).

This can be done easily in the context of M-theory compactified on the
product space of a line interval of length $\pi R_I$ with four circles of
radii $R_1,\cdots$, $R_4$~\cite{op,ap}:
$S^1/Z_2(R_I)\times S^1(R_1)\times T^3(R_2,R_3,R_4)$. One can then
interpret this compactification in various ways by choosing appropriately
one of the radii as that of the eleventh dimension. Considering for
instance $R_I=R_{11}$, one finds the (strongly coupled) heterotic string
compactified on $T^4(R_1,\cdots,R_4)$, while choosing $R_1=R_{11}$, one
finds type IIA compactified on $K3$ of ``squashed" shape
$S^1/Z_2(R_I)\times T^3({\tilde R}_2,{\tilde R}_3,{\tilde R}_4)$, where
the 3 radii ${\tilde R}_i$ will be determined below. In each of the two
cases, one can use the duality relations (\ref{M-het}) to obtain
\be
R_I=\lambda_H l_H=\lambda_{6H}{V_{T^4}^{1/2}\over l_H}
\qquad\qquad R_1=\lambda_{IIA}l_{IIA}
=\lambda_{6IIA}{V_{K3}^{1/2}\over l_{IIA}}\, ,
\ee
while using eqs.(\ref{het-II}) one finds a mapping between the volume of
the internal 4-manifold of one theory and a preferred radius of the
other, in corresponding string units:
\be
{R_I\over l_{IIA}}={V_{T^4}^{1/2}\over l_H^2}\qquad\qquad
{R_1\over l_H}={V_{K3}^{1/2}\over l_{IIA}}\, .
\label{RV}
\ee
The correspondence among the remaining 3 radii can be found, for
instance, by noticing that the S-duality transformations leave invariant
the shape of $T^3$:
\be
{R_i\over R_j}={{\tilde R}_i\over{\tilde R}_j}\qquad\qquad i,j=2,3,4\, ,
\ee
which yields ${\tilde R}_i=l_M^3/(R_jR_k)$ with $i\ne j\ne k\ne i$ and
$l_M^3=\lambda_H l_H^3$. This relation, together with eq.(\ref{RV}),
gives the precise mapping between $T^4$ and $K3$, which completes the
S-duality transformations (\ref{het-II}). We recall that on the type II
side, the four $K3$ directions corresponding to $R_I$ and ${\tilde R}_i$
are transverse to the 5-brane where gauge interactions are localized.

Using the above results, one can now study the possible perturbative type
II descriptions of 4d heterotic compactifications on
$T^4(R_1,\cdots,R_4)\times T^2(R_5,R_6)$ with a certain number $k$ of
large dimensions of common size $R$ and string coupling 
$\lambda_H\sim (R/l_H)^{k/2}\gg 1$. From eq.(\ref{het-II}), the type II
string tension appears as a non-perturbative threshold at energies of the
order of the $T^2$ compactification scale, $l_{II}\sim\sqrt{R_5R_6}$.
Following the steps we used in the context of heterotic -- type I duality,
after T-dualizing the radii which are smaller than the string size, one
can easily show that the $T^2$ directions must be among the $k$ large
dimensions in order to obtain a perturbative type II description. 

It follows that for $k=1$ with, say, $R_6\sim R\gg l_H$, the
type II threshold appears at an intermediate scale
$l_{II}\sim\sqrt{Rl_H}$, together with all 4 directions of $K3$, while
the second, heterotic size, direction of $T^2$ is T-dual (with respect to
$l_{II}$) to $R$: ${\tilde R}_5\equiv l_{II}^2/l_H\sim R$. Thus, one
finds a type IIB description with two large longitudinal dimensions along
the $T^2$ and string coupling of order unity, which is the example
discussed in sections {\it 2.2} and {\it 5.2}.
\begin{scalepic}
{H: $k=1$}{IIB, $\lambda\!\!\sim\!\!1$}{scal01}
\scaleitem{40}{$l_H$, $R_{1,\cdots,4}, R_5$}{1}
\scaleitem{110}{$\sqrt{R}$}{$l_{IIB}$, $K3$}
\scaleitem{180}{$R_6=R$}{$T^2({\tilde R}_5,R_6)$}
\end{scalepic}
For $k\ge 2$, the type II scale
becomes of the order of the compactification scale, $l_{II}\sim R$. For
$k=2$, all directions of $K3\times T^2$ have the type II size, while the
type II  string coupling is infinitesimally small, $\lambda_{II}\sim
l_H/R$, which is the example discussed in section {\it 5.1}. 
\begin{scalepic}
{H: $k=2$}{II,$\lambda\!\!\sim\!\!1\!/\!R$}{scal02}
\scaleitem{40}{$l_H$, $R_{1,\cdots,4}$}{1}
\scaleitem{180}{$R_{5,6}=R$}{$l_{II}$, $K3$, $T^2(R_{5,6})$}
\end{scalepic}
For $k=3$, 
$l_{II}\sim R_{5,6}\sim R$, while the four (transverse) directions of
$K3$ are extra large: $R_I\sim{\tilde R}_i\sim R^{3/2}/l_H$.
\begin{scalepic}
{H: $k=3$}{II, $\lambda\!\!\sim\!\!1$}{scal03}
\scaleitem{40}{$l_H$, $R_{2,3,4}$}{1}
\scaleitem{180}{$R_1=R_{5,6}=R$}{$l_{II}$, $T^2(R_{5,6})$}
\scaleitem{250}{$R^{3/2}$}{$K3$}
\end{scalepic}

For $k=4$, the type II dual theory provides a perturbative description
alternative to the type I$^\prime$ with $n=2$ extra large transverse
dimensions. For $k=5$, there is no perturbative type II description,
while for $k=6$, the heterotic $E_8\times E_8$ theory is described by a
weakly coupled type IIA with all scales of order $R$ apart one $K3$
direction ($R_I$) which is extra large. This is equivalent to type
I$^\prime$ with $n=1$ extra large transverse dimension. Note that this
case was not found from heterotic $SO(32)$ -- type I duality since the
heterotic $SO(32)$ string is equivalent to $E_8\times E_8$ only up to
T-duality, which cannot be performed when $k=6$ and there are no leftover
dimensions of heterotic size. In table \ref{theories}, we summarize the
weakly coupled dual descriptions of the heterotic string with large (TeV)
dimensions, which also provide all possible (perturbative) low-scale
string realizations.
\begin{table}[b]
\caption{Realizations of large dimensions and/or low string scale.
\label{theories}}
\vspace{0.2cm}
\begin{center}
\footnotesize
\begin{tabular}{|c|c|c|c|c|c|}
\hline
Theories & $\parallel$ TeV$^{-1}$ dims & $\perp$ dims & strong gravity & 
string scale \\
\hline
type I/I$^\prime$& $6-n$ & $n\ge 2$ (mm - fm)  & TeV            & TeV \\ 
type IIA & 2     & TeV$^{-1}$          & $10^{19} $ GeV & TeV \\
         & $6-n$ & $2\le n\le 4$ (mm - fm) & TeV            & TeV \\
type IIB & 2     & $10^{11}$ GeV       & $10^{11}$ GeV  & $10^{11}$ GeV \\
\hline
\end{tabular}
\end{center}
\end{table}

\section{Main experimental tests for particle accelerators}

The main predictions of string theories with large volume
compactifications and/or low fundamental scale discussed above follow
from the existence of (i) large {\it longitudinal} dimensions felt by
gauge interactions, (ii) extra large {\it transverse} transverse
dimensions felt only  by gravity that becomes strong at low energies, and
(iii) strings with low tension.

\subsection{Longitudinal dimensions}

They exist generically in all realizations of table
{\ref{theories}, with the exception of type I/I$^\prime$ TeV strings
with six transverse dimensions in the fermi region. Their main
implication is the existence of KK excitations (\ref{pw}) for all
Standard Model gauge bosons and possibly the Higgs~\cite{ia}. They couple
to quarks and leptons which are localized in the compact space, and
generate at low energies four-fermion and higher dimensional effective
operators~\cite{abe}. The coupling of a bulk to two boundary fields
contains a form factor which suppresses exponentially the heavy KK modes
while in the large radius limit it reduces to the 4d gauge coupling.
Thus, the sum over the exchange of KK excitations, which diverges when
the number of extra (longitudinal) dimensions $d_\parallel\ge 2$, is
regulated by the form factor and yields for the strength of 4-fermion
operators $\sim R^2(R/l_s)^{d_\parallel-2}$, in the large radius limit;
for $d_\parallel=2$, there are logarithmic corrections. The current
limits on the size of extra longitudinal dimensions arise from the bounds
of compositeness or from other indirect effects, such as in the Fermi
constant and $Z$-width, and lie in the range of a couple of TeV, if the
string scale is not far from the compactification
scale~\cite{abe,limits}. Otherwise, for $d_\parallel >2$, the limits are
obviously much higher.

The most exciting possibility is of course their discovery through direct
production of KK excitations, for instance in hadron colliders such as 
the Tevatron and LHC, via Drell-Yan  processes~\cite{abq}. The
corresponding KK resonances are narrow with a width-to-mass ratio
$\Gamma/M\sim g^2\sim$  a few per cent, and thus the typical expected
signal is the production of a double (or multiple) resonance for
$d_\parallel=1$ (or $d_\parallel>1$), corresponding to the first KK
mode(s) of the photon and $Z$, very nearly spaced one from the other.
This is depicted in fig.~\ref{fig:KKres} using logarithmic scale. On the
other hand, the  non-observation of deviations from the Standard Model
prediction for the total number of lepton pairs at LHC would translate
into a lower bound of about 7 and 9 TeV, for one and two large
dimensions, respectively.

\begin{figure}
\centerline{\psfig{figure=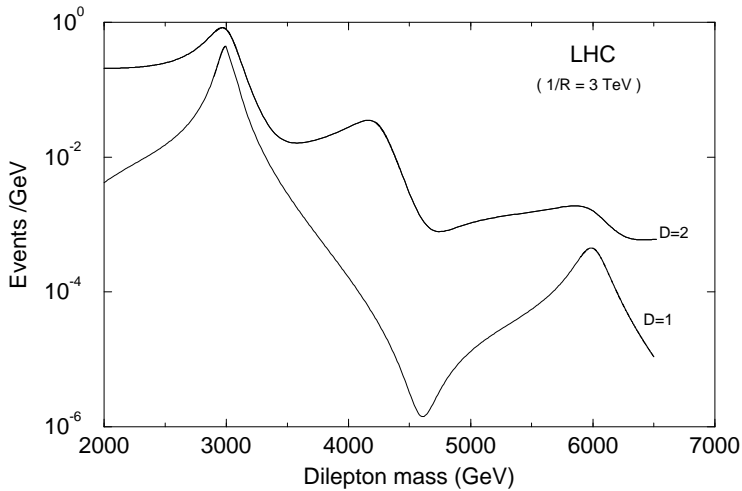,height=7cm}}
\caption{First resonances in the LHC experiment due to a KK state for one
and two extra dimensions at 3 TeV.
\label{fig:KKres}}
\end{figure}

\subsection{Transverse dimensions and low-scale quantum gravity}

They exist generically in all type I/I$^\prime$ realizations of TeV
strings and their size can be as large as a millimeter, which is the
shortest distance up to which gravity has been directly tested
experimentally, as we discussed in section {\it 4.1}. The main
experimental signal in particle accelerators is graviton emission into
the higher-dimensional bulk, leading to jets and missing energy
events~\cite{aadd}.

For illustration, the simplest process is the gluon annihilation into a
graviton which escapes into the extra dimensions. The corresponding
cross-section is given by (in the weak coupling limit)~\cite{aadd}:
\be
\sigma(E)\sim {E^n\over M_I^{n+2}}{\Gamma\left(1-2E^2/M_I^2\right)^2
\over\Gamma\left(1-E^2/M_I^2\right)^4}\, ,
\label{sigma}
\ee
where $E$ is the center of mass energy and $n$ the number of extra large
transverse dimensions. The above expression exhibits 3 kinematic regimes
with different behavior. At high energies $E\gg M_I$, it falls off
exponentially due to the UV softness of strings. At energies of the order
of the string scale, it exhibits a sequence of poles at the position of
Regge resonances. Finally, at low energies $E\ll M_I$, it falls off as a
power $\sigma(E)\sim E^n/M_I^{n+2}$, dictated by the effective higher
dimensional gravity which requires the presence of the
$(4+n)$-dimensional Newton's constant $G_N^{(4+n)}\sim l_I^{n+2}$
from eq.(\ref{GN}).

Since at low energies the effective (higher dimensional) field theory is
valid, one can do reliable computations within it and obtain model
independent predictions that take into account graviton emission in the
bulk~\cite{lowgrav}. It turns out that LHC will be sensitive to
higher-dimensional gravity scales in the range of 3 to 5 TeV, when the
number of transverse dimensions varies from six at the sub-fermi to two
at the sub-millimeter region (where the effect becomes stronger). When
the available energy becomes higher than the gravity scale, gravitational
interactions are strong and particle colliders can become the best probes
for quantum gravity.

\subsection{Low-scale strings}

\noindent The main experimental signal of low-scale strings is 
the production of higher-spin Regge excitations for all Standard Model
particles, with same (internal) quantum numbers and mass-squared
increasing linearly with spin. For instance, the excitations of the gluon
could show up as a series of peaks in jet production at LHC. However, the
corresponding resonances might be very narrow, with a width-to-mass ratio
$\Gamma/M\sim\lambda_I^2\sim$ a few per thousand if 
$v_\parallel\sim 1$ [see eq.(\ref{treei})], and thus difficult to detect.
On the other hand, in the type II realization of TeV strings of table
\ref{theories} using an infinitesimal string coupling
$\lambda_{II}\sim 10^{-14}$, string interactions are extremely
suppressed and there are no observable effects other than KK excitations
of gauge particles, all the way up to (4d) Planckian energies.

\section{Physics}

\subsection{Gauge hierarchy}

In the context of TeV strings, the question of gauge hierarchy, i.e. of
why the Planck mass is much bigger than the weak scale, is translated
into the question of why there are transverse dimensions much larger
than the string scale, or why the string coupling is infinitesimally
small. From eq.(\ref{treei}) in type I/I$^\prime$ strings, the required
hierarchy $R_\perp/l_I$ varies from $10^{15}$ to $10^5$, when the number
of extra dimensions in the bulk varies from $n=2$ to $n=6$, respectively,
while in type II strings with no large dimensions, the required value of
the coupling $\lambda_{II}$ is $10^{-14}$.

Besides the hard dynamical question on the origin of the hierarchy, there
is a technical aspect, which consists of understanding its stability
against possible large quantum corrections. This is precisely the problem
that supersymmetry solves, softly broken at the TeV scale. In our
context, this problem can be studied by examining the limit of
decompactification $R_\perp/l_I\to\infty$, or of vanishing coupling
$\lambda_{II}\to 0$. As we will see below, in general this limit does not
exist, implying for instance that the transverse space does not decouple
in the decompactification limit~\cite{ab}. The reason is that the UV
cutoff of the effective field theory on the brane is not always the
string scale but the winding scale $R_\perp M_I^2$, dual to the large
transverse dimensions, which is much larger than the type I/I$^\prime$
string tension~\cite{cb}. This can happen when the number of transverse
dimensions is less or equal to two, or more generally when there is
effective propagation of gravity in one or two transverse dimensions.

The source of this divergence is the emission of (massless) closed string
tadpoles in the bulk, which can be	attached to any physical amplitude
involving open string fields living on our world-brane. Thus, the
potential divergence is a string infrared effect but, from the point of
view of the brane theory, it looks as a UV correction that modifies its
low-energy effective couplings. The contribution of these {\it local}
tadpoles $\cal T$ to the world-brane amplitudes can be estimated easily
as follows~\cite{ab}:
\be
{\cal T}\sim\  {1\over V_\perp}\  \sum_{\vert {p}_\perp\vert < M_{s} }\
{1\over p_\perp^2}\  F({\vec p_\perp})\, ,
\label{tadpole}
\ee
where $V_\perp={R_\perp}^{d_\perp}$ is the volume of the transverse
space, ${\vec p}_\perp =(m_1/{R_\perp}$$\cdots$
${m_{d_\perp}/{R_\perp}})$ is the transverse momentum carried away by the
massless closed string, and the sum is restricted to transverse distances
$l_\perp$ large compared to the string length 
$l_\perp\sim 1/p_\perp\gg l_s$. $F({\vec p_\perp})$ are the local
tadpoles, Fourier-transformed to momentum space, arising from the
distribution of the D-branes and the orientifolds that act as classical
point-like sources in the transverse space. Consistency of the theory
requires the global tadpole cancellation condition $F(0)=0$, implying the
vanishing of the total charge (D-branes versus
orientifolds)~\cite{strings}. In the simplest toroidal compactification
(with vanishing antisymmetric tensor), this fixes the number of D-branes
to 32 and tadpoles take the generic form:
\be
F({\vec p_\perp})\sim\left( 32\prod_{i=1}^{d_\perp}{1+(-)^{m_i}\over 2}
-2\sum_{a=1}^{16}{\rm cos}({\vec p_\perp}{\vec x_a})\right)\, ,
\ee
where the orientifolds are located at the corners of the cell 
$[0, \pi R_\perp]^{d_\perp}$, and $\pm{\vec x_a}$ are the transverse
positions of the 32 D-branes, which correspond to Wilson lines of the
T-dual picture.

For generic positions of the D-branes, the tadpole contribution
(\ref{tadpole}) has the following behavior in the decompactification
limit:
\be
{\cal T}\sim \cases{ O(R_\perp)\ &for \ \ \ $d_\perp=1$\cr
O(\ln R_\perp)\ \ \ &for \ \ \ $d_\perp=2$\cr
O(1) \ &for\ \ \ \ $d_\perp>2$\cr}\quad ,
\ee
which is dictated by the large-distance behavior of the two-point
function in the $d_\perp$-dimensional transverse space. It follows that
when there is one dimension much larger than the others ($d_\perp=1$),
there are in general large linear corrections yielding through
eq.(\ref{treei}) quadratic UV divergences regulated by the 4d Planck
mass, $R_\perp\sim M_P^2/M_I^3$. In general, one expects such large
corrections to occur in particular in gauge kinetic terms, that drive the
theory rapidly to a strong coupling singularity and, thus, forbid the
size of the transverse space to become much larger than the string length.
This is precisely the phenomenon we studied in section 3, for the 11th
dimension of M-theory compactified on $S^1/Z_2``\times"$CY.

The conclusion is that the technical aspect of gauge hierarchy is solved
without the need of supersymmetry in the following two cases~\cite{ab}.
(i) In special models in which tadpoles cancel locally in the transverse
space. In the one-dimensional case ($d_\perp=1$), this happens when
D-branes are equally distributed at the two fixed points of the
orientifold, generalizing the condition of symmetric embedding in
M-theory compactifications discussed in section 3. (ii) When $d_\perp\ge
2$. The limiting case $d_\perp=2$ is particularly attractive because it
allows the effective couplings of the brane theory to depend
logarithmically on the size of the transverse space, or equivalently on
$M_P$, exactly as in the case of softly broken supersymmetry. Moreover,
similarly to renormalizable quantum field theories, the logarithmic
divergences can be absorbed into a finite number of parameters, that
correspond to the values of bulk fields at the (transverse) position of
our world-brane which determine all effective couplings of the brane
theory. In addition, the renormalization group resummation is replaced by
the {\it classical} equations of motion of the effective 2d supergravity
in the transverse space, with higher-derivative terms being ignored
because the variations of fields are logarithmic. As a result, the case
of $d_\perp=2$ leaves open the possibility of dynamically determining the
hierarchy, by minimizing an effective potential on our world-brane that
depends logarithmically on the size of transverse space~\cite{gv}. This is
again in analogy to the inverse hierarchy idea in supersymmetric field
theories.

It turns out that low-scale type II theories with infinitesimal string
coupling share many common properties with type I$^\prime$ when
$d_\perp=2$~\cite{ap}. In fact, the limit of vanishing coupling does not
exist due to subtleties related to the singular character of the
compactification manifold and to the non perturbative origin of gauge
symmetries. In general, there are corrections depending logarithmically
on the string coupling, similarly to the case of type I$^\prime$ strings
with 2 transverse dimensions.

\subsection{Unification}

One of the motivations for supersymmetry comes from the apparent
unification of gauge couplings discussed in section {\it 2.1}. It is then
important to study this issue in the context of the new framework of
low-scale strings. One possibility is to use power-law running that may
accelerate unification in an energy region where the theory becomes
higher dimensional~\cite{ddg}. Within the effective field theory, the
summation over the KK modes above the compactification scale and below
some energy scale
$E\gg R^{-1}$ yields:
\be
{1\over g^2_a(E)}={1\over g^2_a(R^{-1})}-{b_a^{SM}\over 2\pi}\ln(ER)-
{b_a^{KK}\over 2\pi}\left\{ c_d\left[ (ER)^d-1\right]-\ln(ER)\right\}\, ,
\label{powerev}
\ee
where $c_d=\pi^{d/2}/d\Gamma(1+d/2)$ for $d$ extra (longitudinal)
dimensions. The first logarithmic term corresponds to the usual 4d
running controlled by the Standard Model beta-functions $b_a^{SM}$, while
the next term is the contribution of the KK tower dominated by the
power-like dependence $(ER)^d$ associated to the effective multiplicity
of KK modes and controlled by the corresponding beta-functions
$b_a^{KK}$.

In supersymmetric theories, the KK excitations have at least $N=2$
extended supersymmetry obtained by standard dimensional reduction of the
higher-dimensional theory. Assuming the MSSM particle content below the
compactification scale, its minimal $N=2$ extension requires gauge boson
excitations to form $N=2$ vector multiplets, containing for every spin-1
a Dirac fermion and 2 real scalars, while higgs and matter multiplets do
not apriori need to have excitations if they belong to boundary
(twisted-like) states. It was observed however that if higgs excitations
form, level by level, one $N=2$ hypermultiplet\footnote{This is the case
when for instance one higgs doublet comes from the bulk and the other
from the boundary.} (containing 1 Dirac fermion and 4 scalars), the
unification of gauge couplings is approximately maintained for any value
of $R$, but it arises very rapidly above the compactification scale due
to the power evolution (\ref{powerev})~\cite{ddg}. For instance, when
$d=1$ and $R^{-1}\simeq 1$ TeV, the gauge couplings meet around 50 TeV
within 2\%, while the five-dimensional coupling $g_a(ER)^{1/2}$ remains
perturbative. The main disadvantage of this approach is that the result
is very sensitive (power-like) to the initial conditions and thus to
string threshold corrections, in contrast to the usual unification based
on logarithmic evolution.

This scenario requires obviously that the string scale is low and,
therefore, should be analyzed in the context of type I/I$^\prime$
superstring theory. It turns out that in supersymmetric vacua string loop
corrections to gauge couplings diverge at most quadratically with the
radius, even if there are more than two large dimensions ($d>2$).
Moreover, in type I/I$^\prime$ theory the quadratic terms are included in
the tree-level expression of the couplings, leaving only the possibility
of linearly divergent corrections when $d=1$~\cite{cb,abd}. 

On the other hand, following the analysis of the previous subsection 
{\it 7.1}, there is an alternative possibility to obtain large threshold
corrections when the effective transverse dimensionality of the bulk is
$d_\perp\le 2$. In particular, when $d_\perp=2$, there are logarithmic
corrections that could restore the ``old" unification picture with a GUT
scale given by the winding scale, which for millimeter-size dimensions
has the correct order of magnitude~\cite{cb,ab,admr}. In this way, the
running due to the large desert in energies is replaced by an effective
running due to the ``large desert" in transverse distances from our
world-brane. However, an explicit computation of string threshold
corrections in $N=1$ orientifolds shows that both the linear and
logarithmic contributions are controlled by the corresponding $N=2$
$\beta$-functions and, thus, are model dependent~\cite{abd}.

Indeed, the one-loop corrected gauge couplings in $N=1$ orientifolds are
given by the following expression:
\be
{1\over g^2_a(\mu)}={1\over g^2}+s_a m+
{b_a\over 4\pi}\ln{M_I^2\over\mu^2}-\sum_{i=1}^3{b_{a,i}^{N=2}\over 4\pi}
\left\{\ln T_i +f(U_i)\right\}\, ,
\label{thresholds}
\ee
where the first two terms in the r.h.s. correspond to the tree-level
(disk) contribution and the remaining ones are the one-loop (genus-1)
corrections. Here, we assumed that all gauge group factors correspond to
the same type of D-branes, so that gauge couplings are the same to lowest
order (given by $g$). $m$ denotes a combination of the twisted moduli,
whose VEVs blow-up the orbifold singularities and allow the transition to
smooth (Calabi-Yau) manifolds. However, in all known examples, these VEVs
are fixed to $m=0$ from the vanishing of the D-terms of anomalous
$U(1)$'s.

As expected, the one-loop corrections contain an infrared divergence,
regulated by the low-energy scale $\mu$, that produces the usual 4d
running controlled by the $N=1$ beta-functions $b_a$. The last sum
displays the string threshold corrections that receive contributions only
from $N=2$ sectors, controlled by the corresponding $N=2$
beta-functions $b_{a,i}^{N=2}$. They depend on the geometric moduli
$T_i$ and $U_i$, parameterizing the size and complex structure of the
three internal compactification planes. In the simplest case of a
rectangular torus of radii $R_1$ and $R_2$, $T=R_1R_2/l_I^2$ and
$U=R_1/R_2$. The function $f(U)=\ln\left({\rm Re}U|\eta(iU)|^4\right)$
with $\eta$ the Dedekind-eta function; for large $U$, $f(U)$ grows
linearly with $U$.  Thus, from expression (\ref{thresholds}), it follows
that when $R_1\sim R_2$, there are logarithmic corrections 
$\sim\ln(R_1/l_I)$, while when $R_1>R_2$, the corrections grow linearly
as $R_1/R_2$. Note that in both cases, the corrections are proportional
to the $N=2$ $\beta$-functions. Obviously, unification based on
logarithmic evolution requires the two (transverse) radii to be much
larger than the string length, while power-low unification can happen
either when there is one longitudinal dimension a bit larger than the
string scale  ($R_1/R_2\sim R_\parallel/l_I$ keeping $\lambda_I<1$), or
when one transverse direction is bigger than the rest of the bulk.

\subsection{Supersymmetry breaking}

Following the discussion of subsection {\it 7.1}, TeV scale strings offer
a solution to the technical (at least) aspect of gauge hierarchy without
the need of supersymmetry, provided there is no effective propagation of
bulk fields in a single transverse dimension, or else closed string
tadpoles should cancel locally. It is then natural to ask the question
whether there is any motivation leftover for supersymmetry or not. This
comes from the problem of the cosmological constant~\cite{aadd}. 

In fact, in a non-supersymmetric string theory, the bulk energy density
behaves generically as $\Lambda_{\rm bulk}\sim M_s^{4+n}$, where $n$ is
the number of transverse dimensions much larger than the string length.
In the type I/I$^\prime$ context, this induces a cosmological constant on
our world-brane which is enhanced by the volume of the transverse space
$V_\perp\sim R_\perp^n$. When expressed in terms of the 4d parameters
using the type I/I$^\prime$ mass-relation (\ref{treei}), it is translated
to a quadratically dependent contribution on the Planck mass:
\be
\Lambda_{\rm brane}\sim M_I^{4+n}R_\perp^n\sim M_I^2 M_P^2\, ,
\label{lambda}
\ee
where we used $s=I$. This contribution is in fact the analogue of the
quadratic divergent term Str${\cal M}^2$ in softly broken supersymmetric
theories, with $M_I$ playing the role of the supersymmetry breaking
scale. 

The brane energy density (\ref{lambda}) is far above the (low) string
scale $M_I$ and in general destabilizes the hierarchy that one tries to
enforce. One way out is to resort to special models with broken
supersymmetry and vanishing or exponentially small cosmological
constant~\cite{ks}. Alternatively, one could conceive a different
scenario, with supersymmetry broken primordially on our world-brane
maximally, i.e. at the string scale which is of order of a few TeV. In
this case the brane cosmological constant would be, by construction,
${\cal O}(M_I^4)$, while the bulk would only be affected by
gravitationally suppressed radiative corrections and thus would be almost
supersymmetric~\cite{aadd,ads}. In particular, one would expect the
gravitino and other soft masses in the bulk to be  extremely small
$O(M_I^2/M_P)$. In this case, the cosmological constant induced in the
bulk would be
\be
\Lambda_{\rm bulk}\sim M_I^4/R_\perp^n\sim M_I^{6+n}/M_P^2\, ,
\label{lambdasmall}
\ee
i.e. of order (10 MeV)$^6$ for $n=2$ and $M_I\simeq 1$ TeV.
The scenario of brane supersymmetry breaking is also required in models
with a string scale at intermediate energies $\sim 10^{11}$ GeV (or
lower), discussed in section {\it 4.1}. It can occur for instance on a
brane distant from our world and is then mediated to us by gravitational
(or gauge) interactions.\\
\begin{center}
{\it Brane supersymmetry breaking}
\end{center}
\vskip .2cm
\noindent In the absence of gravity, brane supersymmetry breaking can
occur in a non-BPS system of rotating or intersecting D-branes. Since
brane rotations correspond to turning on background magnetic fields, they
can be easily generalized in the presence of gravity, in the context of
type I string theory~\cite{ba}. The main problems of this approach are
the generic appearance of tadpoles, the presence of tachyons and the lack
of gaugino masses. Stable non-BPS configurations of intersecting branes
have been studied more recently~\cite{sen}, while their implementation in
type I theory was achieved only very recently~\cite{ads}.

The simplest examples are based on orientifold projections of type IIB,
in which some of the orientifold 5-planes have opposite charge, requiring
an open string sector living on anti-D5 branes in order to cancel the RR
(Ramond-Ramond) charge. As a result, supersymmetry is broken on the
intersection of D9 and anti-D5 branes that coincides with the world
volume of the latter. The simplest construction of this type is a 
$T^4/Z_2$ orientifold with a flip of the $\Omega$-projection (world-sheet
parity) in the twisted orbifold sector. It turns out that several
orientifold models, where tadpole conditions do not admit naive
supersymmetric solutions, can be defined by introducing
non-supersymmetric open sector containing anti-D-branes. A typical
example of this type is the ordinary $Z_2\times Z_2$ orientifold with
discrete torsion. 

The resulting models are chiral, anomaly-free, with vanishing RR tadpoles
and no tachyons in their spectrum~\cite{ads}. Supersymmetry is broken at
the string scale on a collection of anti-D5 branes while, to lowest
order, the closed string bulk and the other branes are supersymmetric. In
higher orders, supersymmetry breaking is of course mediated to the
remaining sectors, but is suppressed by the size of the transverse space
or by the distance from the brane where supersymmetry breaking primarily
occurred. The models contain in general uncancelled NS (Neveu-Schwarz)
tadpoles reflecting the existence of a tree-level potential for the NS
moduli, which is localized on the (non-supersymmetric) world volume of
the anti-D5 branes.

As a result, this scenario implies the absence of supersymmetry on our
world-brane but its presence in the bulk, a millimeter away! The bulk
supergravity is needed to guarantee the stability of gauge hierarchy
against large gravitational quantum radiative corrections.\\
\begin{center}
{\it Low-scale type II models}
\end{center}
\vskip .2cm
\noindent Note that the above destabilization problem does not exist in
low-scale type II vacua with no large dimensions but infinitesimal string
coupling, since in this case the (one-loop) vacuum energy behaves as
$\Lambda\sim M_{II}^4$. On the other hand, in type IIB vacua with two
large (TeV$^{-1}$) longitudinal dimensions and string scale at
intermediate energies, discussed in section {\it 5.2}, supersymmetry
breaking could arise for instance by Scherk-Schwarz compactification at a
scale 
$m_{susy}\sim R^{-1}\simeq M_{IIB}^2/M_P$~\cite{kp,ia}.\footnote{This
mechanism can also be used to break supersymmetry in type I/I$^\prime$
models along a compact dimension of size not very different from the
string length~\cite{ssopen}.} This is in line with the original
motivation of large dimensions in the context of the heterotic (dual)
theory, discussed in section {\it 2.2}, and leads to a vacuum energy that
behaves as $\Lambda\sim 1/R^4$, up to logarithmic
corrections~\cite{ia,iadd}. This behavior is due to the extreme softness
of the mechanism of supersymmetry breaking realized through a change of
boundary conditions, similarly to the effects of finite temperature upon
the identification $T\equiv R^{-1}$. Indeed, the summation over the KK
excitations amounts to inserting the Boltzmann factors $e^{-E/T}$ to all
thermodynamic quantities --or equivalently to the soft breaking terms-- 
that suppresses exponentially their UV behavior.

The extreme softness of supersymmetry breaking by compactification
implies a particular spectroscopy of superparticles that differs
drastically from other scenarios~\cite{ia,adpq}. In the simplest case,
supersymmetry breaking generates a universal tree-level mass for gaugini,
while scalar masses vanish to lowest order. Moreover, the latter are
insensitive to the UV cutoff at one loop, and thus squarks and leptons
are naturally an order of magnitude lighter than gaugini. On the other
hand, if the Higgs scalar lives in the bulk of the extra (TeV)
dimension(s), a heavy higgsino mass is automatically generated and there
is no $\mu$-problem.  These models offer also the possibility of
determining the hierarchy by minimizing the effective potential which
acquires at higher loops logarithmic corrections in $R$.

\section{Gravity modification and sub-millimeter forces}

Besides the spectacular experimental predictions in particle
accelerators, string theories with large volume compactifications and/or
low string scale predict also possible modifications of gravitation in
the sub-millimeter range, which can be tested in ``tabletop" experiments
that measure gravity at short distances. There are two categories of such
predictions:\hfil\\  
(i) Deviations from the Newton's law $1/r^2$ behavior to $1/r^{2+n}$, for
$n$ extra large transverse dimensions, which can be observable for $n=2$ 
dimensions of sub-millimeter size. This case is particularly attractive
on theoretical grounds because of the logarithmic sensitivity of Standard
Model couplings on the size of transverse space, but also for
phenomenological reasons since the effects in particle colliders are
maximally enhanced~\cite{lowgrav}. Notice also the coincidence of this
scale with the possible value of the cosmological constant in the
universe that recent observations seem to support.\hfil\\
(ii) New scalar forces in the sub-millimeter range, motivated by the
problem of supersymmetry breaking discussed in section {\it 7.3}, and
mediated by light scalar fields $\varphi$ with
masses~\cite{fkz,iadd,aadd,ads}:
\be
m_{\varphi}\simeq{m_{susy}^2\over M_P}\simeq 
10^{-4}-10^{-2}\ {\rm eV} \, ,
\label{msusy}
\ee
for a supersymmetry breaking scale $m_{susy}\simeq 1-10$ TeV. These
correspond to Compton wavelengths in the range of 1 mm to 10 $\mu$m.
$m_{susy}$ can be either the KK scale $1/R$ if supersymmetry is broken by
compactification~\cite{iadd}, or the string scale if it is broken
``maximally" on our world-brane~\cite{aadd,ads}. A model independent
scalar mediating the force is the radius modulus (in Planck units)
\be
\varphi\equiv\ln R\, ,
\label{varphi}
\ee
with $R$ the radius of the longitudinal or transverse dimension(s),
respectively. In the former case, the result (\ref{msusy}) follows from
the behavior of the vacuum energy density $\Lambda \sim 1/R^4$ for large
$R$ (up to logarithmic corrections). In the latter case, supersymmetry is
broken primarily on the brane only, and thus its transmission to the bulk
is gravitationally suppressed, leading to masses (\ref{msusy}).

The coupling of these light scalars to nuclei can be computed since it
arises dominantly through the radius dependence of $\Lambda_{\rm QCD}$,
or equivalently of the QCD gauge coupling. More precisely, the coupling
$\alpha_\phi$ of the radius modulus (\ref{varphi}) relative to gravity
is~\cite{iadd}:
\be
\alpha_\varphi = {1\over m_N}{\partial m_N\over\partial\varphi}=
{\partial\ln\Lambda_{\rm QCD}\over\partial\ln R}= 
-{2\pi\over b_{\rm QCD}}{\partial\over\partial\ln R}\alpha_{\rm QCD}\, ,
\label{coupling}
\ee
with $m_N$ the nucleon mass and $b_{\rm QCD}$ the one-loop QCD
beta-function coefficient. In the case where supersymmetry is broken via
Scherk-Schwarz compactification, the couplings are radius-independent in
the supersymmetric limit and thus 
$\partial_\varphi\alpha_{\rm QCD}=(b_{\rm QCD}-b_{\rm SQCD})/2\pi$, with
$b_{\rm SQCD}$ the supersymmetric beta-function. Using $b_{\rm SQCD}=3$
and $b_{\rm QCD}=7$, one finds that the force between two pieces of
matter mediated by the radius modulus is
$\alpha_\varphi^2=(4/7)^2\simeq 1/3$ times the force of gravity. On the
other hand, in the case where supersymmetry is broken primordially on our
world-brane at the string scale while it is almost unbroken the bulk, the
force (\ref{coupling}) is again comparable to gravity in theories with
logarithmic sensitivity on the size of transverse space, i.e. when there
is effective propagation of gravity in $d_\perp=2$ transverse dimensions.
The resulting forces can therefore be within reach of upcoming
experiments~\cite{price}.

In principle there can be other light moduli which couple with even
larger strengths. For example the dilaton $\phi$, whose VEV determines
the (logarithm of the) string coupling constant, if it does not acquire
large mass from some dynamical supersymmetric mechanism, can lead to a
force of strength 2000 times bigger than gravity~\cite{tvdil}.

\begin{figure}
\centerline{\psfig{figure=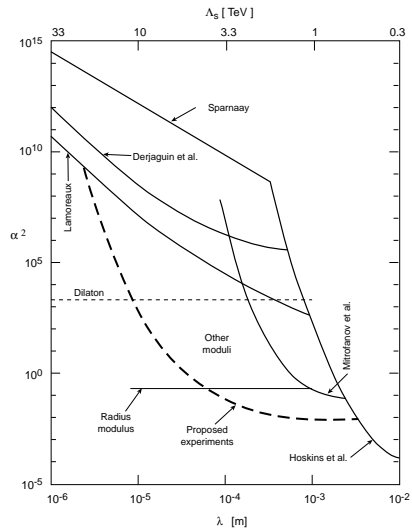,height=7cm}}
\caption{Strength of the modulus force relative to gravity ($\alpha^2$)
versus  its Compton wavelength ($\lambda$).
\label{fig:forces}}
\end{figure}

In fig.~\ref{fig:forces} we depict the actual information from previous,
present and upcoming experiments~\cite{price}. The vertical axis is the
strength, $\alpha^2$, of the force relative to gravity; the horizontal
axis is the Compton wavelength of the exchanged particle; the upper scale
shows the corresponding value of the supersymmetry breaking scale (large
radius or string scale) in TeV. The solid lines indicate the present
limits from the experiments indicated. The excluded regions lie above
these solid lines. Measuring gravitational strength forces at such short  
distances is quite challenging. The most important background is the Van
der Walls force which becomes equal to the gravitational force between
two atoms when they are about 100 microns apart. Since the Van der Walls
force falls off as the 7th power of the distance, it rapidly becomes
negligible compared to gravity at distances exceeding 100 $\mu$m. The
dashed thick line gives the expected sensitivity of the present and
upcoming experiments, which will improve the actual limits by roughly two
orders of magnitude and --at the very least-- they will, for the first
time, measure gravity to a precision of 1\% at distances of $\sim$ 100
$\mu$m.

\section*{Acknowledgments}
Research supported in part by the EEC under the TMR contract
ERBFMRX-CT96-0090.


\section*{References}
\begin{thebibliography}{99}
\bibitem{strings} M. Green, J. Schwarz and E. Witten, {\em Superstring
Theory}, Cambridge University Press, 1987; J. Polchinski, {\em String
Theory}, Cambridge University Press, Cambridge, 1998.

\bibitem{sao} For recent reviews, see A. Sen, hep-th/9802051,
I. Antoniadis and G. Ovarlez, hep-th/9906108.

\bibitem{ia} I. Antoniadis, \Journal{\PLB}{246}{377}{1990}.

\bibitem{dkl} L. Dixon, V. Kaplunovsky and J. Louis,
\Journal{\NPB}{355}{649}{1991}; I. Antoniadis, K. Narain and T. Taylor,
\Journal{\PLB}{267}{37}{1991}.

\bibitem{ablt} I. Antoniadis, C. Bachas, D. Lewellen and T. Tomaras,
\Journal{\PLB}{207}{441}{1988}.

\bibitem{kp} C. Kounnas and M. Porrati, \Journal{\NPB}{310}{355}{1988};
S. Ferrara, C. Kounnas, M. Porrati and F. Zwirner,
\Journal{\NPB}{318}{75}{1989}; E. Kiritsis and C. Kounnas,
\Journal{\NPB}{503}{117}{1997}.

\bibitem{tv} T. Taylor and G. Veneziano, \Journal{\PLB}{212}{147}{1988}.

\bibitem{ap} I. Antoniadis and B. Pioline, \Journal{\NPB}{550}{41}{1999},
hep-th/9902055.

\bibitem{w95} E. Witten, 
Proceedings of Strings 95, hep-th/9507121; A.~Strominger,
\Journal{\PLB}{383}{44}{1996}, hep-th/9512059;

\bibitem{sei} N. Seiberg, \Journal{\PLB}{390}{169}{1997}, hep-th/9609161.

\bibitem{w} E. Witten, \Journal{\NPB}{471}{135}{1996}, hep-th/9602070.

\bibitem{l} J.D. Lykken, \Journal{\PRD}{54}{3693}{1996}, hep-th/9603133.

\bibitem{add} N. Arkani-Hamed, S. Dimopoulos and G. Dvali, 
\Journal{\PLB}{429}{263}{1998}, hep-ph/9803315.

\bibitem{aadd} I. Antoniadis, N. Arkani-Hamed, S. Dimopoulos and G. Dvali, 
\Journal{\PLB}{436}{263}{1998}, hep-ph/9804398. 

\bibitem{ab} I. Antoniadis and C. Bachas, 
\Journal{\PLB}{450}{83}{1999}, hep-th/9812093.

\bibitem{hw} P. Ho\v{r}ava and E. Witten, 
\Journal{\NPB}{460}{506}{1996}, hep-th/9510209.

\bibitem{ckm} E. Caceres, V.S. Kaplunovsky and I.M. Mandelberg,
\Journal{\NPB}{493}{73}{1997}, hep-th/9606036.

\bibitem{pw} J. Polchinski and E. Witten,
\Journal{\NPB}{460}{525}{1996}, hep-th/9510169.

\bibitem{st} G. Shiu and S.-H.H. Tye, 
\Journal{\PRD}{58}{106007}{1998}, hep-th/9805157; Z. Kakushadze and
S.-H.H. Tye, \Journal{\NPB}{548}{180}{1999}, hep-th/9809147;
L.E. Ib\'a\~nez, C. Mu\~noz and S. Rigolin, hep-ph/9812397.

\bibitem{add2} N. Arkani-Hamed, S. Dimopoulos and G. Dvali,
\Journal{\PRD}{59}{086004}{1999}, hep-ph/9807344;
S. Nussinov and R. Shrock, \Journal{\PRD}{59}{105002}{1999},
hep-ph/9811323; S. Cullen and M. Perelstein,
\Journal{\PRL}{83}{268}{1999}, hep-ph/9903422; L.J. Hall and D. Smith,
hep-ph/9904267.

\bibitem{price} See for instance: J.C. Long, H.W. Chan and J.C. Price,
\Journal{\NPB}{539}{23}{1999}, hep-ph/9805217.

\bibitem{li} L.E. Ib\'a\~nez, Proceedings of Strings 99.

\bibitem{int} K. Benakli, hep-ph/9809582; C.P. Burgess, L.E. Ib\'a\~nez
and F. Quevedo, \Journal{\PLB}{447}{257}{1999}.

\bibitem{aq} I. Antoniadis and M. Quir{\'o}s, 
\Journal{\PLB}{392}{61}{1997}, hep-th/9609209.

\bibitem{ht} C.M. Hull and P.K. Townsend, 
\Journal{\NPB}{438}{109}{1995}, hep-th/9410167 and
\Journal{\NPB}{451}{525}{1995}, hep-th/9505073;
E. Witten, 
\Journal{\NPB}{443}{85}{1995}, hep-th/9503124.

\bibitem{pm} For a recent review, see P. Mayr, hep-th/9904115.

\bibitem{kv} S. Katz and C. Vafa, 
\Journal{\NPB}{497}{146}{1997}, hep-th/9606086; for a recent
review, see P. Mayr, 
{\em Fortsch. Phys.} {\bf 47}, 39 (1998), hep-th/9807096. 

\bibitem{op} For a recent review, see N.A. Obers and B. Pioline,
{\em Phys. Rep.} {\bf 318}, 113 (1999), hep-th/9809039.

\bibitem{abe} I. Antoniadis and K. Benakli,
\Journal{\PLB}{326}{69}{1994}, hep-th/9310151. 

\bibitem{limits} V.A. Kostelecky and S. Samuel,
\Journal{\PLB}{270}{21}{1991}; P. Nath and M. Yamaguchi, hep-ph/9902323
and hep-ph/9903298; W.J. Marciano, hep-ph/9902332 and hep-ph/9903451; M.
Masip and A. Pomarol, hep-ph/9902467. 

\bibitem{abq} I. Antoniadis, K. Benakli and M. Quir{\'o}s,
\Journal{\PLB}{331}{313}{1994}, hep-ph/9403290 and 
\Journal{\PLB}{460}{176}{1999}, hep-ph/9905311;
P. Nath, Y. Yamada and M. Yamaguchi, hep-ph/9905415;
T.G. Rizzo and J.D. Wells, hep-ph/9906234; A. Strumia, hep-ph/9906266.

\bibitem{lowgrav} See for instance: G.F. Giudice, R. Rattazzi and J.D.
Wells, \Journal{\NPB}{544}{3}{1999}, hep-th/9811292;  E.A. Mirabelli, M.
Perelstein and M.E. Peskin, \Journal{\PRL}{82}{2236}{1999},
hep-ph/9811337; T. Han, J. Lykken and R.-J. Zhang,
\Journal{\PRD}{59}{105006}{1999}, hep-ph/9811350; J.L. Hewett,
\Journal{\PRL}{82}{4765}{1999}, hep-ph/9811356; T.G. Rizzo,
\Journal{\PRD}{59}{115010}{1999}, hep-ph/9901209, 
\Journal{\PRD}{60}{075001}{1999}, hep-ph/9903475
and hep-ph/9904380; P. Mathews, S. Raychaudhuri and K. Sridhar,
\Journal{\PLB}{450}{343}{1999}, hep-ph/9811501 and hep-ph/9904232;  
G. Shiu, R. Shrock and S.-H.H. Tye, \Journal{\PLB}{458}{274}{1999},
hep-ph/9904262; E. Halyo, hep-ph/9904432; M. Besancon, hep-ph/9909364.

\bibitem{cb} C. Bachas, 
{\em JHEP} {\bf 9811}, 23 (1998), hep-ph/9807415.

\bibitem{gv} See also G. Dvali, \Journal{\PLB}{459}{489}{1999},
hep-ph/9905204.

\bibitem{ddg} K.R. Dienes, E. Dudas and T. Gherghetta, 
\Journal{\PLB}{436}{55}{1998}, hep-ph/9803466 and
\Journal{\NPB}{537}{47}{1999}, hep-ph/9806292; D. Ghilencea and G.G.
Ross, \Journal{\PLB}{442}{165}{1998}; Z. Kakushadze, 
\Journal{\NPB}{548}{205}{1999}, hep-th/9811193; 
A. Delgado and M. Quir\'os, hep-ph/9903400; 
P. Frampton and A. Rasin, \Journal{\PLB}{460}{313}{1999}, hep-ph/9903479;
A. P\'erez-Lorenzana and R.N. Mohapatra, hep-ph/9904504;
Z. Kakushadze and T.R. Taylor, hep-th/9905137.

\bibitem{abd} I. Antoniadis, C. Bachas and E. Dudas, hep-th/9906039.

\bibitem{admr} N. Arkani-Hamed, S. Dimopoulos and J. March-Russell,
hep-th/9908146.

\bibitem{ks} S. Kachru and E. Silverstein, {\em JHEP} {\bf 11}, 1 (1998),
hep-th/9810129; J. Harvey, \Journal{\PRD}{59}{26002}{1999}; R.
Blumenhagen and L. G{\"o}rlich, hep-th/9812158; C. Angelantonj, I.
Antoniadis and K. Foerger, \Journal{\NPB}{555}{116}{1999},
hep-th/9904092.

\bibitem{ads} I. Antoniadis, E. Dudas and A. Sagnotti, hep-th/9908023;
G. Aldazabal and A.M. Uranga, hep-th/9908072.

\bibitem{ba} C. Bachas, hep-th/9503030;
J.G. Russo and A.A. Tseytlin, \Journal{\NPB}{461}{131}{1996}.

\bibitem{sen} For recent reviews, see A. Sen, hep-th/9904207; 
A. Lerda and R. Russo, hep-th/9905006.

\bibitem{ssopen} I. Antoniadis, E. Dudas and A. Sagnotti,
\Journal{\NPB}{544}{469}{1999}; I. Antoniadis, G. D'Appollonio, E. Dudas
and A. Sagnotti, \Journal{\NPB}{553}{133}{99}, hep-th/9812118 and
hep-th/9907184.

\bibitem{iadd} I. Antoniadis, S. Dimopoulos and G. Dvali,
\Journal{\NPB}{516}{70}{1998}, hep-ph/9710204.

\bibitem{adpq} I. Antoniadis, C. Mu\~noz and M. Quir\'os,
\Journal{\NPB}{397}{515}{1993}; A. Pomarol and M. Quir\'os,
\Journal{\PLB}{438}{225}{1998}, hep-ph/9806263;
I. Antoniadis, S. Dimopoulos, A. Pomarol and M. Quir\'os,
\Journal{\NPB}{544}{503}{1999}, hep-ph/9810410; A. Delgado, A. Pomarol
and M. Quir\'os, hep-ph/9812489.

\bibitem{fkz} S. Ferrara, C. Kounnas and F. Zwirner,
\Journal{\NPB}{429}{589}{1994}.

\bibitem{tvdil} T.R. Taylor and G. Veneziano,
\Journal{\PLB}{213}{450}{1988}.


\end{thebibliography}
\end{document}